\documentclass{aastex63}
\usepackage{comment}
\usepackage{lineno}
\usepackage{array,multirow}
\shorttitle{Dynamical ages of three clusters from the dynamical clock}
\shortauthors{Ferraro et al.}

\graphicspath{{./}{figures/}}

\begin{document}

\title{Empirical measurement of the dynamical ages of three globular clusters 
 and some considerations on the use of the dynamical clock\footnote{Based on observations collected at the Hubble Space
  Telescope, under proposal GO12517 (PI: Ferraro), GO13410 (PI:
  Pallanca), GO15232 (PI: Ferraro).}}

\correspondingauthor{Francesco R. Ferraro}
\email{francesco.ferraro3@unibo.it}

\author[0000-0002-2165-8528]{Francesco R. Ferraro}
\affil{Dipartimento di Fisica e Astronomia, Universit\`a di Bologna, Via Gobetti 93/2 I-40129 Bologna, Italy}
\affil{INAF-Osservatorio di Astrofisica e Scienze dello Spazio di Bologna, Via Gobetti 93/3 I-40129 Bologna, Italy}

\author[0000-0001-5613-4938]{Barbara Lanzoni}
\affil{Dipartimento di Fisica e Astronomia, Universit\`a di Bologna, Via Gobetti 93/2 I-40129 Bologna, Italy}
\affil{INAF-Osservatorio di Astrofisica e Scienze dello Spazio di Bologna, Via Gobetti 93/3 I-40129 Bologna, Italy}
 
\author[0000-0003-2742-6872]{Enrico Vesperini}
\affil{Department of Astronomy, Indiana University, Bloomington, IN, 47401, USA}

\author[0000-0002-5038-3914]{Mario Cadelano}
\affil{Dipartimento di Fisica e Astronomia, Universit\`a di Bologna, Via Gobetti 93/2 I-40129 Bologna, Italy}
\affil{INAF-Osservatorio di Astrofisica e Scienze dello Spazio di Bologna, Via Gobetti 93/3 I-40129 Bologna, Italy}

\author[0000-0002-3730-9664]{Dan Deras}
\affil{Dipartimento di Fisica e Astronomia, Universit\`a di Bologna, Via Gobetti 93/2 I-40129 Bologna, Italy}
 
\author[0000-0002-7104-2107]{Cristina Pallanca}
\affil{Dipartimento di Fisica e Astronomia, Universit\`a di Bologna, Via Gobetti 93/2 I-40129 Bologna, Italy}
\affil{INAF-Osservatorio di Astrofisica e Scienze dello Spazio di Bologna, Via Gobetti 93/3 I-40129 Bologna, Italy}





\begin{abstract}
We have used the ``dynamical clock'' to measure the level of dynamical
evolution reached by three Galactic globular clusters (namely, NGC
3201, NGC 6316 and NGC 6440). This is an empirical method that
quantifies the level of central segregation of blue stragglers stars
(BSSs) within the cluster half-mass radius by means of the $A^+_{rh}$
parameter, defined as the area enclosed between the cumulative radial
distribution of BSSs and that of a lighter population. The total
sample with homogeneous determinations of $A^+_{rh}$ now counts a
gran-total of 59 clusters: 52 old GCs in the Milky Way (including the
three investigated here), 5 old clusters in the Large Magellanic
Cloud, and 2 young systems in the Small Magellanic Cloud.  The three
objects studied here nicely nest into the correlation between
$A^+_{rh}$ and the central relaxation time defined by the previous
sample, thus proving and consolidating the use of the dynamical clock
as an excellent tracer of the stage of star cluster dynamical
evolution in different galactic environments.  Finally, we
  discuss the advantages of using the dynamical clock as an indicator
  of star cluster dynamical ages, compared to the present-day central
  relaxation time.
\end{abstract}



\section{Introduction}
\label{sec:intro}
Globular Clusters (GCs) are the prototypes of ``collisional stellar
systems'' in the Universe.  Recurrent gravitational interactions among
their constituent stars favor continuous kinetic-energy exchanges
driving the system toward energy equipartition and inducing severe
perturbations to the stellar orbits.  Heavy stars tend to
progressively sink toward the central region of the cluster (due to
dynamical friction), while low-mass stars migrate outward and can even
escape from the (mass-segregated) system.  The redistribution of
kinetic energy tends to completely erase the initial kinematical and
structural conditions, bringing the cluster toward a (quasi)
thermodynamically relaxed state \citep[see,
  e.g.,][]{trenti+13,bianchini+16} in a characteristic timescale that
can be significantly shorter than its age.  The relaxation time
depends in a very complex way on initial and the local conditions,
thus differing from cluster to cluster and, within the same system,
from high- to low-density regions (e.g., \citealt{meylan+97}).  Thus,
even clusters formed at the same epoch (i.e., with the same
\emph{chronological age}) are expected to show different levels of
dynamical evolution (namely, different \emph{dynamical ages}).  The
macroscopic manifestation of internal changes induced by the dynamical
evolution of the system is a progressive contraction of the central
regions (in particular, of the core radius, $r_c$) and a corresponding
increase of the central density ($\rho_0$) virtually up to infinity,
in a runaway process that is called ``core collapse'' and is thought
to be halted by the formation and hardening of binary systems
\citep[e.g.][]{meylan+97}. However, an opposite behavior, where the
core radius progressively expands with time due to the heating effect
of a retained population of stellar mass black holes, has been
advocated \citep{mackey+08} to explain the size-age conundrum observed
in the Magellanic Clouds, where young star clusters are all compact,
while the old ones show both small and large core sizes
(\citealp{mackey+03a, mackey+03b}).  This indicates how difficult and
uncertain is to estimate the dynamical age of stellar systems solely
from measuring their structural parameters, and clearly calls for
additional methods providing a more direct empirical measure of the
effects of the various processes driving the dynamical evolution of
star clusters.  Among these, the measure of the stellar mass function,
orbital anisotropy, velocity dispersion profile at different radial
distances or for different stellar mass groups (e.g., \citealt{bm03,
  bianchini+16, bianchini+18, tiongco+16, webb+17, cadelano+20a,
  beccari+22}) is still quite challenging for most GCs (e.g,
\citealt{libralato+18,libralato+19, cohen+21}), while the study of
special classes of heavy objects seems to be particularly
promising. Indeed, the intense dynamical activity in GC interiors is
thought to boost the formation of stellar exotica, like blue straggler
stars (BSSs) and binaries containing heavily degenerate objects, like
black holes (BHs) and neutron stars (NSs).  Hence, from one side the
frequency and the properties of these exotica are expected to depend
on the dynamical stage of the system, on the other side, the
observational properties of this special class of objects can be used
to get information on the internal dynamics of GCs \citep[see,
  e.g.,][]{pooley+03, ransom+05, ferraro+09, ferraro+18a, ferraro+19,
  dalessandro+13, verbunt+14, cadelano+17a, cadelano+18, cadelano+19,
  cadelano+20b, prager+17, beccari+19}.

Among the variety of exotica populating GC cores, BSSs are surely the
most abundant and the easiest to distinguish from normal stars in a
colour-magnitude diagram (CMD), where they populate a sort of
extension of the cluster main-sequence (MS) toward brighter magnitudes
and bluer colors than the MS turn-off (MS-TO) point
\citep[e.g.,][]{sandage53, ferraro+92, ferraro+93, ferraro+97,
  ferraro+99a, ferraro+03, ferraro+06a, piotto+04, lanzoni+07a,
  lanzoni+07b, lanzoni+07c, leigh+07, moretti+08, dalessandro+08,
  beccari+11, simunovic+16}). Their ``anomalous'' location in the CMD
suggests that BSSs are hydrogen-burning stars more massive than the
others. The origin of such massive objects in stellar systems with no
gas available for star formation requires the action of some
mass-enhancement process. Two main BSS formation channels have been
identified so far: mass-transfer in binary systems \citep{mccrea64},
and stellar mergers resulting from direct collisions \citep{hills+74,
  sills+05}). There is also growing observational evidence
\citep{shara+97, gilliland+98, ferraro+06b, fiorentino+14, raso+19}
confirming that they are indeed significantly heavier ($m_{\rm
  BSS}=1.2 M_\odot$) than the average cluster population ($\langle m
\rangle =0.3 M_\odot$). Hence, these objects represent powerful
gravitational probes of processes that characterize the dynamical
evolution of star clusters \citep[e.g.][]{ferraro+09, ferraro+12,
  ferraro+18a, ferraro+19, lanzoni+16, dresbach+22}.  Indeed, the
signature of mass segregation and dynamical friction is expected to
remain imprinted in some BSS observational properties and, in fact,
the radial distribution of these stars with respect to normal
(lighter) cluster populations used as reference (REF) has been found
to be a powerful (and fully empirical) indicator of the level of
dynamical evolution reached by the host system, yielding to the
definition of the so-called \emph{``dynamical clock''}
\citep{ferraro+12}.  To solve complications related to the choice of
radial binning, the original definition of the dynamical clock was
later refined with the introduction of the $A^+_{rh}$ parameter
\citep{alessandrini+16, lanzoni+16, ferraro+18a, ferraro+20}, which is
the area enclosed between the cumulative radial distribution of BSSs
and that of REF stars measured within the cluster half-mass radius
($r_h$). Hence, by construction, the $A^+_{rh}$ parameter quantifies
the level of central segregation of BSSs with respect to the REF
population, and it is expected to progressively increase with the
dynamical ageing of the host stellar system due to the more rapid
sedimentation of heavier stars with respect to less massive
objects. Large values of $A^+_{rh}$ are therefore expected for GCs in
late stages of their dynamical evolution, while small values (down to
zero) are predicted in dynamically-young systems, where dynamical
friction has not been effective yet in segregating BSSs toward the
center.  The $A^+_{rh}$ parameter therefore provides a direct,
empirical measure of the central segregation of the heaviest
observable stars within a cluster, as set by the combination of all
the known (and still unknown) internal and external processes that
drive mass segregation. In addition, it is defined within the
  half-mass radius, which is the physical scale-length expected to
  vary the least during cluster dynamical evolution (see, e.g., Figure
  4 in \citealp{bhat+22}).
 
Indeed, a tight relation between $A^+_{rh}$ and the central relaxation
time ($t_{rc}$)
has been discovered for the Galactic GCs investigated so far (all
sharing approximately the same chronological age), thus providing
their direct ranking in terms of dynamical age \citep{lanzoni+16,
  ferraro+18a}.  The application of the dynamical clock to a sample of
old star clusters in the Large Magellanic Cloud (LMC) has demonstrated
that they follow the same relation defined by the Galactic ones
\citep{ferraro+19}.  This has also contributed to clarify the origin
of the size-age conundrum, showing that the different core radii
measured for the old GCs in the sample can be naturally explained by
different dynamical ages (the dynamically older systems having smaller
$r_c$ than the younger clusters), with no need of a retained
population of black holes.  \citet{ferraro+19} also pointed out that
the distribution of the LMC clusters in the $r_c-$age diagram in no
way can be interpreted as an evolutionary sequence because of the
strong difference in mass between the young and the old clusters,
being less and more massive than $10^5 M_\odot$, respectively.
Finally, \citealt{dresbach+22} recently demonstrated that the
correlation found for the old GCs in the Galaxy and the LMC also holds
for NGC 339 and NGC 419, two intermediate-age GCs in the Small
Magellanic Cloud (SMC).

\begin{figure}[ht!]
\centering
\includegraphics[width=18cm, height=8.5cm]{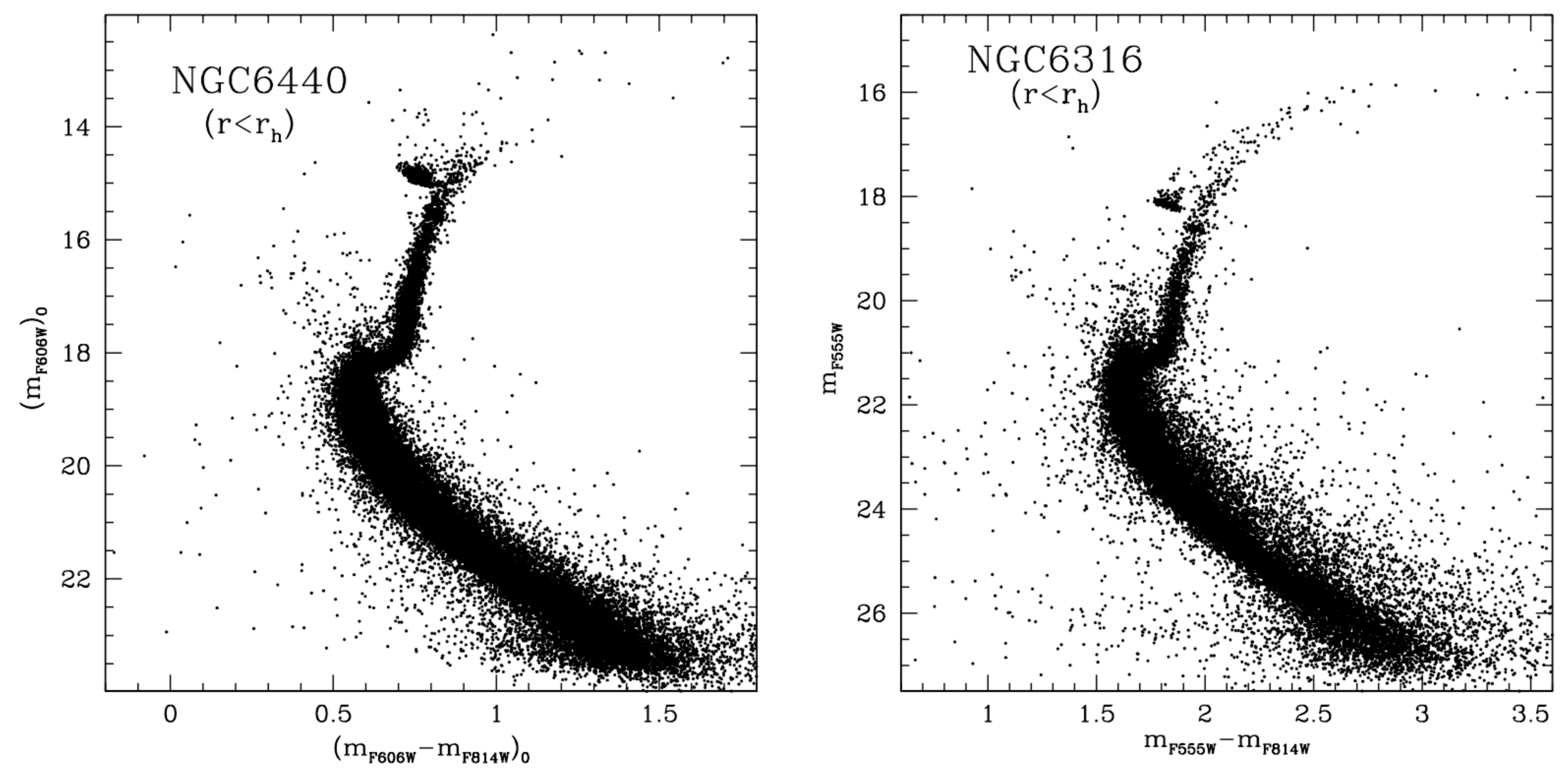}
\centering
\caption{CMDs of NGC 6440 (left) and NGC 6316 (right) within one
  half-mass radius from the center of each cluster: $r_h=50.2\arcsec$
  \citep{pallanca+21} and $r_h=40\arcsec$ \citep{deras+23},
  respectively.  }
\label{cdm_6440_6316}
\end{figure}

As part of the project ``Cosmic-Lab'', which is aimed at using star
clusters in the local Universe as cosmic laboratories to study the
complex interplay between the dynamical evolution of stellar systems
and the photometric, kinematical and chemical properties of their
stellar populations (see, e.g., \citealt{lanzoni+10, lanzoni+13,
  lanzoni+16, lanzoni+19, miocchi+13, cadelano+17a, ferraro+18a,
  ferraro+18b, ferraro+19, ferraro+20, ferraro+21, raso+19,raso+20,
  pallanca+21}), here we study the BSS population and measure the
$A^+_{rh}$ parameter in three additional Galactic GCs, namely, NGC
3201, NGC 6316 and NGC 6440. The paper is organized as follows. In
Section \ref{sec:obs} we summarize the observations and the adopted
data reduction procedures.  In Section \ref{sec:analysis} we discuss
the selection of the BSS samples and the measure of $A^+_{rh}$. The
discussion of the results and the comparison with other Galactic and
extra-Galactic clusters is presented in Section \ref{sec:discussion}.

\section{Observations and data reduction}
\label{sec:obs}
To characterize the BSS distribution in each of the investigated
clusters, we took advantage of recent photometric studies performed by
our group: \citet{pallanca+19, pallanca+21} for NGC 6440,
\citet{deras+23} for NGC 6316, \citet{ferraro+18a} and Lanzoni et
al. (2023, in preparation) for NGC 3201. In Table \ref{tab1} we list
the main characteristics of the clusters under investigation and the
respective reference paper where the photometric dataset and the
photometric analysis is described in detail. In the following we
shortly summarize the relevant information about the adopted datasets
and the results of the photometric analyses.
 
\begin{figure}[ht!]
\centering
\includegraphics[width=18cm, height=6.4cm]{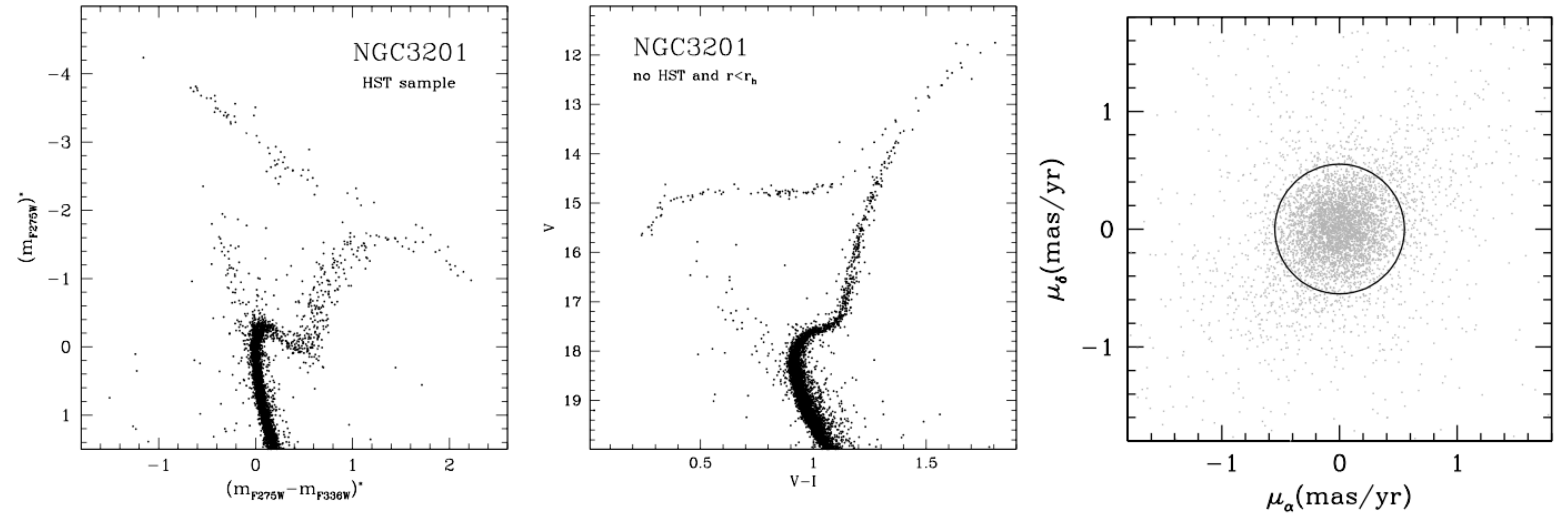}
\centering
\caption{{\it Left:} UV CMD of NGC 3201 obtained from HST/WFC3 data
  sampling the innermost $\sim 100\arcsec$ from the center. The
  observed values of $m_{\rm F275}$ and $(m_{\rm F336W}- m_{\rm
    F336W})$ have been shifted to locate the MS-TO at zero magnitude
  and color: $(m_{\rm F275})^*_{\rm MS-TO}=0$ and $(m_{\rm F336W}-
  m_{\rm F336W})^*_{\rm MS-TO}=0$ \citep[see][]{ferraro+18a}.  {\it
    Center:} Differential reddening corrected optical CMD of NGC 3201
  obtained from the Stetson photometric catalog \citep{stetson_cat}
  for the stars located beyond the HST/WFC3 field of view and within
  $r_h=272\arcsec$ (Lanzoni et al. 2023).  Only member stars, selected
  from Gaia PMs as shown in the right panel, are plotted. {\it Right:}
  Vector-point diagram for the stars with $V<19.5$ shown in the
  central panel.  The circle has a radius of $2\sigma$, with $\sigma$
  being the average PM dispersion in the two dimensions. The stars
  included within the circle are considered cluster members, and
  plotted in the central panel.}
\label{cmd_3201}
\end{figure}

\begin{itemize}
\item{\bf NGC 6440 $-$} The photometric dataset consists in a series
  of deep images of the cluster central regions acquired with the
  Hubble Space Telescope (HST) Wide Field Camera 3 (WFC3) in different
  filters (especially the F606W and F814W).  To sample the external
  portion of the system, \citet{pallanca+21} used a combination of
  ground-based data acquired with the FOcal Reducer/low dispersion
  Spectrograph2 (FORS2) mounted at the ESO Very Large Telescope, and
  the Pan-STARRS catalog.  The photometric analysis has been carried
  out by applying the point-spread function (PSF) technique to each
  exposure.  We first modeled the PSF using dozens of bright,
  isolated, and non saturated stars. The model was then applied to all
  the sources detected at a given level (3-5 $\sigma$) above the
  background.  We used the DAOPHOT/ALLFRAME package \citet{daophot,
    allframe} and followed the standard procedure adopted in many
  previous papers \citep[see, e.g., ][]{cadelano+17a, cadelano+19} a
  result, a photometric catalog listing the frame coordinates and the
  instrumental magnitudes measured in all the filters for all the
  detected sources is then obtained. Finally, geometric distortions
  have been corrected following the prescriptions of
  \citealt{bellini+11}, and the frame coordinates have been reported
  into right ascension and declination as defined in the World
  Coordinate System by using a sample of stars in common with the
  publicly available Gaia DR2 catalog \citealt{gaia16a,gaia16b} . The
  resulting astrometric accuracy turns out to be smaller than $\sim
  0.1\arcsec$. In the case of the HST data, the instrumental
  magnitudes have been calibrated to the VEGAMAG system by using the
  reference photometric zeropoints reported on the WFC3 web site.  The
  available dataset allowed \citet{pallanca+19, pallanca+21} to
  correct the HST CMD for differential reddening effects, thus
  providing updated estimates of the cluster age, distance, and
  absolute reddening.  By taking advantage of multi-epoch HST
  observations, individual proper motions (PMs) have been determined
  and used to decontaminate the cluster population within $100\arcsec$
  from the centre from field star interlopers.  A new identity card of
  NGC 6440 was thus obtained, with all the structural parameters (as
  the core and half-mass radii, the King concentration parameter, the
  center of gravity, etc.) being re-determined from its resolved star
  density profile. Since the main aim of the present paper is to
  determine the stage of dynamical evolution of each system through
  the measure the $A^+_{rh}$ parameter, in the following we focus our
  attention on the cluster population included within
  $r_{h}=50.2\arcsec$ \citep{pallanca+21}.  The left panel of Figure
  \ref{cdm_6440_6316} shows the reddening corrected and PM-selected
  CMD of NGC 6440 within one half-mass radius.
  
\item{\bf NGC 6316 $-$} Deep optical observations obtained with the
  HST WFC3 in the F555W and the F814W filters have been recently used
  to analyze the stellar population and the structure of this poorly
  investigated GC in the Galactic bulge \citep{deras+23}.  The data
  reduction procedure is very similar to that described above for NGC
  6440, and all the details can be found in \citealt{deras+23}.  Also
  in this case, a high-resolution exctinction map in the direction of
  the system was determined and used to correct the CMD for the
  effects of differential reddening.  The final CMD extends down to
  $m_{\rm F555W}=27$, reaching more than 5 magnitudes below the MS-TO,
  and it clearly delineates the presence of a metal-rich stellar
  population, with a well-defined red clump and a well visible red
  giant branch bump \citep{fusipecci+90, ferraro+99b, ferraro+00,
    valenti+04, valenti+07}.  To sample the entire radial extension of
  the cluster, the HST data (which cover the innermost $\sim
  120\arcsec$) have been complemented with the Gaia DR3 catalog
  \citealt{gaia3}. The cluster structural parameters have then been
  determined from the \citet{king+66} model fit to the resolved star
  density profile.  The differential reddening corrected CMD of NGC
  6316 within one half mass radius ($r_h=40\arcsec$) from the center
  is shown in the right panel of Figure \ref{cdm_6440_6316}.

\item{\bf NGC 3201 -} The central portion of the cluster has been
  sampled in the context of the HST UV Legacy Survey of GCs
  \citep{piotto+15} through deep WFC3 observations in the F275W and
  F336W filters.  The photometric analysis is described in
  \citet{ferraro+18a, nardiello+18}. Briefly, for each image we
  obtained an optimal array of PSFs to properly take into account both
  the spatial and the temporal PSF variations. To extract the
  photometric catalogs from each individual exposure by using the
  adopted arrays of PSFs, we used the software described in
  \citet{anderson+08} properly adapted to WFC3 images.  As above, the
  stellar centroids have been corrected for geometric distortion
  \citep{bellini+11} and transformed to the absolute coordinate
  system, while the instrumental magnitudes have been calibrated to
  the VEGAMAG system by using the reference photometric zero-points
  reported on the WFC3 web page. The memebership probability based on
  HST PMs has also been determined for each star.  This dataset has
  been complemented by a multi-band photometric catalog from the
  Stetson database \citep{stetson_cat}, which samples clustercentric
  distances out to $\sim 25\arcmin$. This has been corrected for
  differential reddening by following the procedure described in
  \citet[][see also \citealt{dalessandro+18}]{cadelano+20c} and then
  cross-correlated with the Gaia DR3 catalog to select cluster members
  on the basis of the measured PMs (see the right panel of Figure
  \ref{cmd_3201}).  The resolved star density profile of NGC 3201 has
  been recently obtained from these data, and the cluster structural
  parameters have been determined from the best-fit King model to the
  observed distribution (Lanzoni et al. 2023). Figure \ref{cmd_3201}
  shows the PM-selected CMD of the two photometric catalogs used to
  sample the cluster population within one half-mass radius ($r_h=
  272\arcsec$).
\end{itemize}

 \begin{figure}[ht!]
\centering \includegraphics[width=18cm, height=8.2cm]{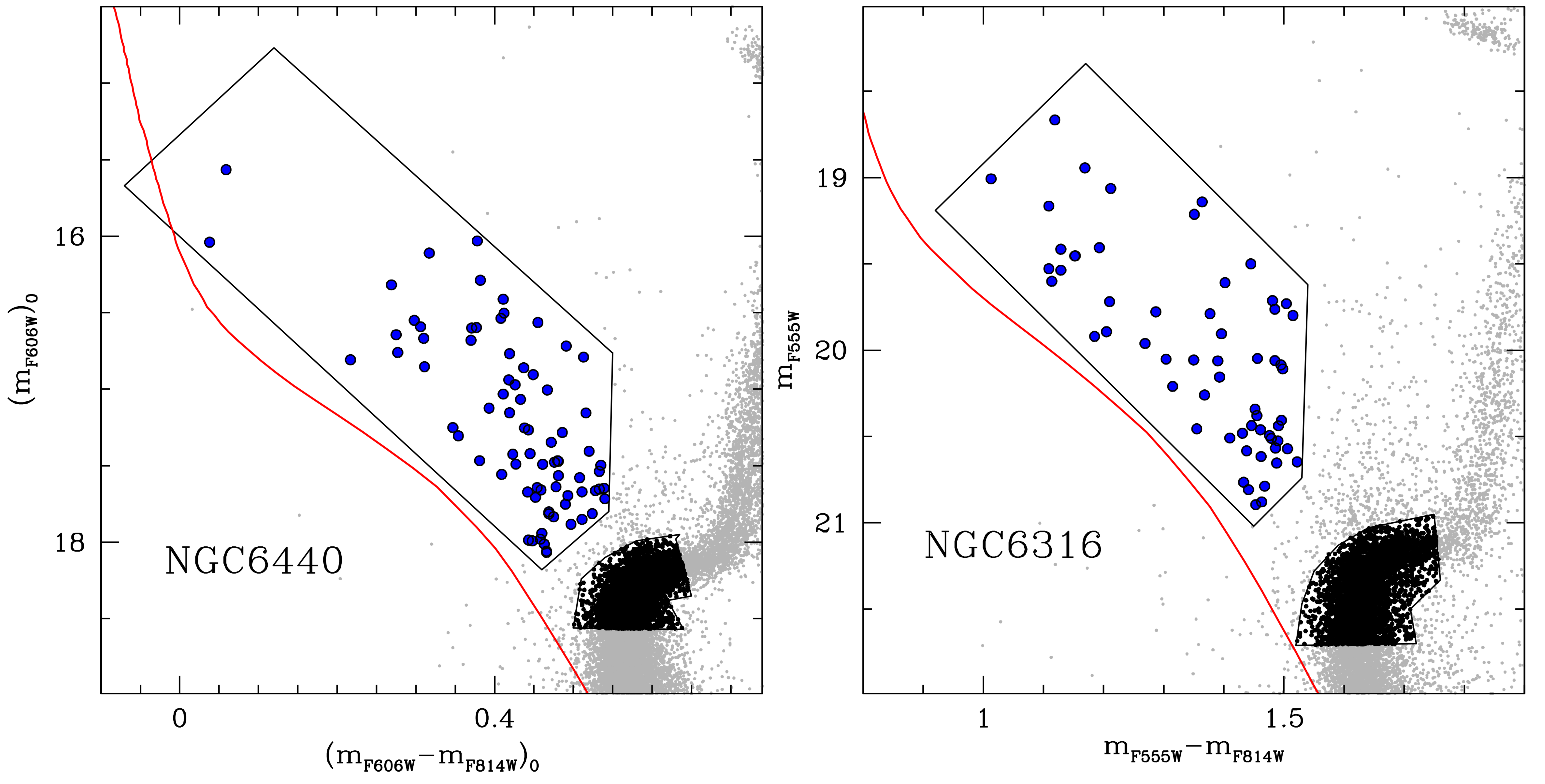}
\centering
\caption{BSS and REF populations (blue and black circles,
  respectively) selected in NGC 6440 (left) and NGC 6316 (right)
  within the cluster half-mass radius. The red lines are BaSTI
    isochrones \citealt{pietrinferni+06,pietrinferni+21} with a very young age (40 Myr) representing
    the ZAMS of each system.}
\label{bss1640}
\end{figure}

 \begin{figure}[ht!]
\centering
\includegraphics[width=18cm, height=8.2cm]{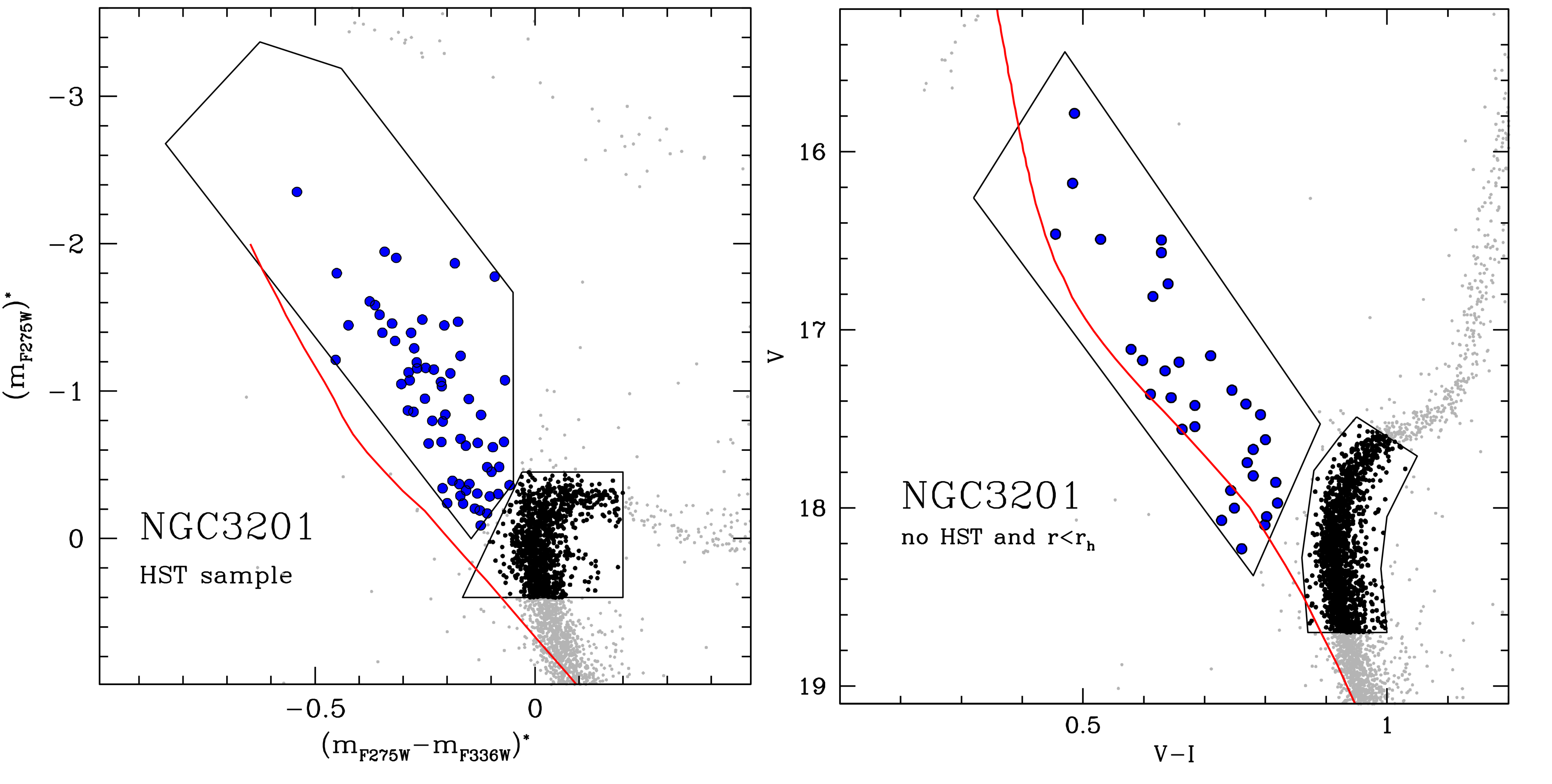}
\centering
\caption{As in Figure \ref{bss1640}, but for NGC 3201 in the UV and
  optical CDMs (left and right panels, respectively) sampling the
  region included within the cluster half-mass radius
  ($r_h=272\arcsec$).}
\label{bss32}
\end{figure}

 \begin{figure}[ht!]
\centering
\includegraphics[width=12cm, height=12cm]{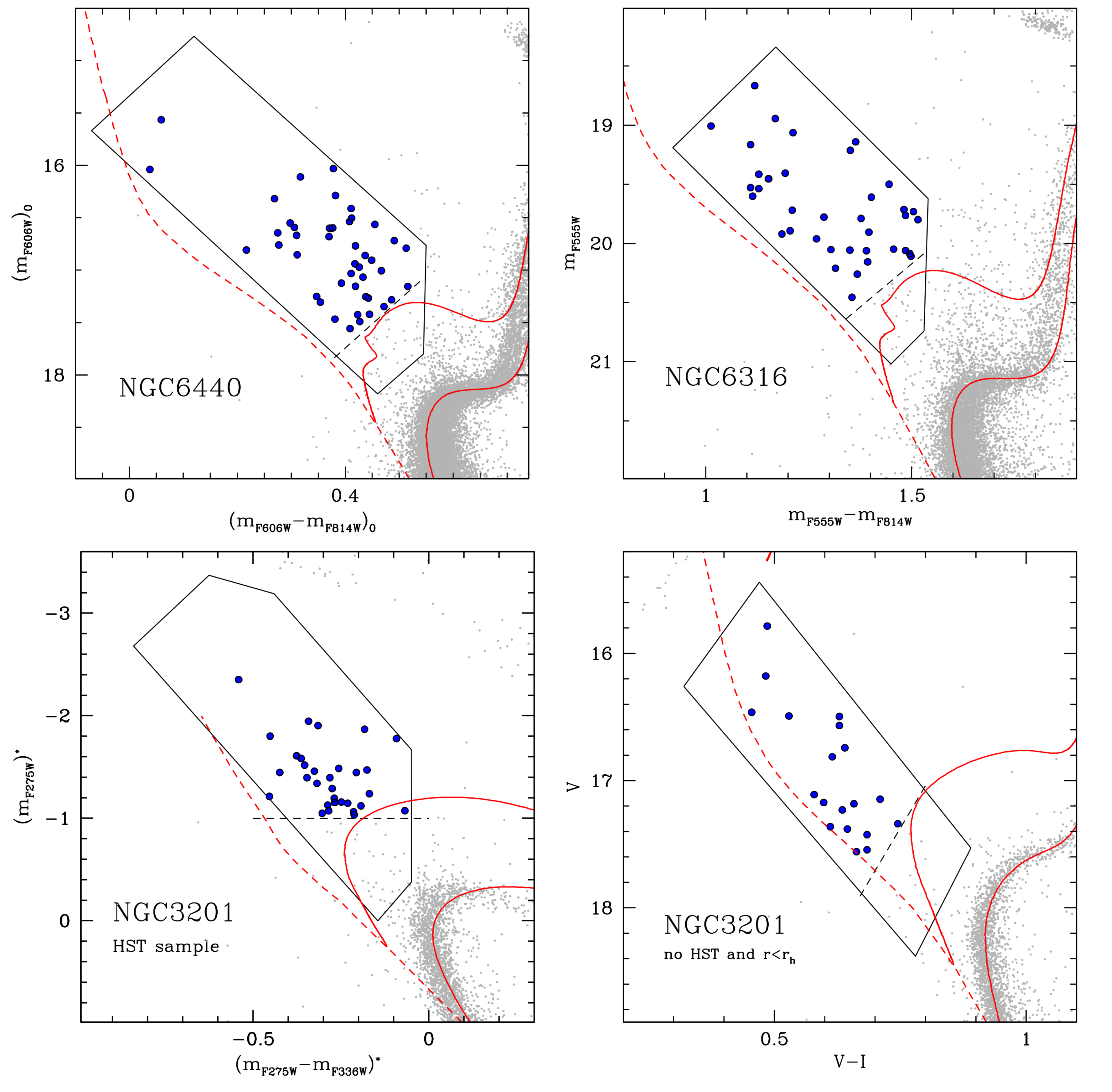}
\centering
\caption{CMDs of the three program clustes with the bright-BSS
  population (used to compute $A^+_{rh}$) highlighted with large blue
  circles.  In each panel, the rightmost red line is the 12.5 Gyr old
  BaSTI isochrone \citealt{pietrinferni+06,pietrinferni+21} of appropriate metallicity that well
  reproduces the cluster MS-TO region, and that has been used to
  estimate the MS-TO mass. The leftmost red line in each panel is the
  BaSTI evolutionary track corresponding to a stellar mass $0.2
  M_\odot$ larger than the MS-TO mass. This has been used to draw the
  dashed lines in the optical CMDs, above which the bright-BSS samples
  (large blue circles) have been selected.  The BSS selection in the
  UV CMD (bottom-left panel) has been performed as in
  \citet{ferraro+18a}, at $(m_{\rm F275})^*<-1$, and it very well
  corresponds to the procedure adopted at optical
  wavelengths.}
\label{bbss}
\end{figure}

 \begin{figure}[ht!]
\centering
\includegraphics[width=18cm, height=6.4cm]{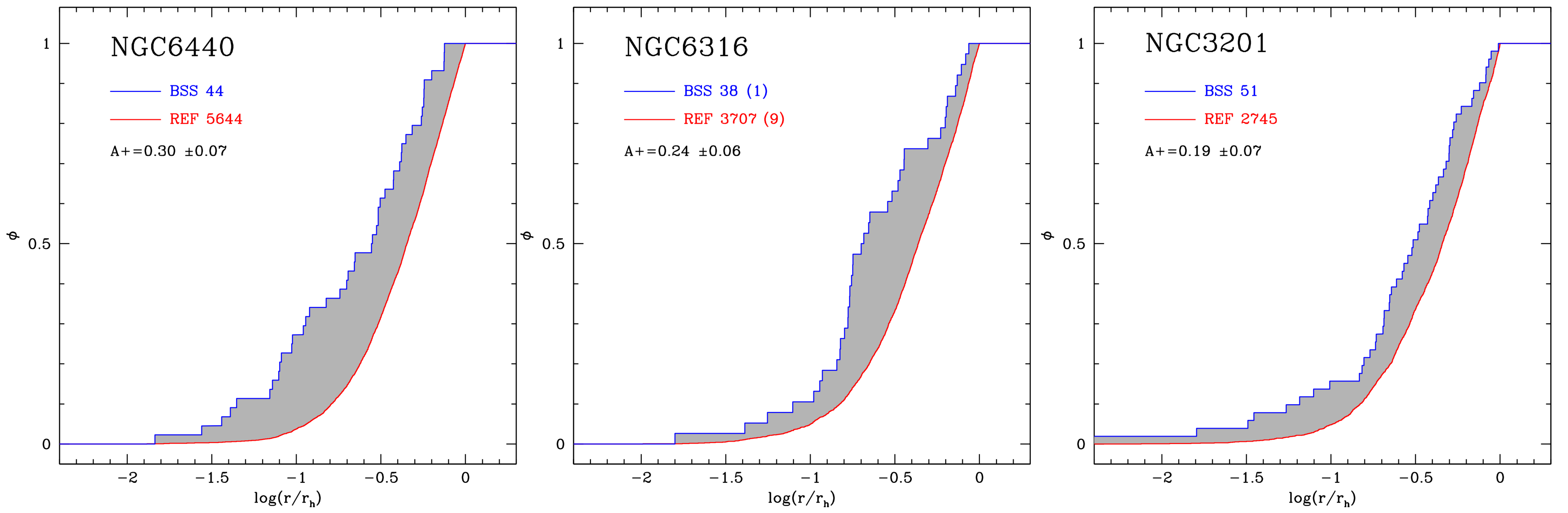}
\centering
\caption{Cumulative radial distributions of BSSs (blue lines) and REF
  stars (red lines) in the three Galactic GCs discussed in this
  paper. Only stars within one half-mass radius have been considered
  and the cumulative radial distributions are therefore normalized to
  unity at $r_h$. The size of the area between the two curves (shaded
  in grey) corresponds to the value of $A^+_{rh}$ labelled in each
  panel. The number of BSS and REF stars selected in each cluster
    is labelled each respective panel. The number of contaminating
    field stars is quoted in parenthesis for NGC6316.}
\label{apiu}
\end{figure}

\section{Analysis}
\label{sec:analysis} 
\subsection{Population selection} 
\label{sec:selection}
The first step for the measure of the $A^+_{rh}$ parameter is the
selection of the BSS population in each cluster.  As discussed in many
previous papers, BSSs are easily recognizable in any CMD independently
of the specific filter combination, because they always populate the
region that is bluer and brighter than the cluster MS-TO point.  To
select the BSS population, we followed the procedure adopted in many
previous studies over the last 30 years (see, e.g.,
\citealt{ferraro+92, ferraro+93, ferraro+99a, ferraro+01, ferraro+03,
  ferraro+06a}). In general, we define a selection box as a
five/six-side polygon aimed at optimizing the separation of BSSs from
cluster stars populating the other evolutionary sequences.  Typically,
the polygon consists of two main lines defining a diagonal
strip. Ideally the line that defines the lower diagonal boundary runs
close and nearly parallel to the zero-age MS (ZAMS) in that specific
filter combination. The upper diagonal boundary is essentially
parallel to this first line and it is set to include the bulk of the
BSS population.  The definition of the bottom edge separating the BSS
population from MS-TO and sub-giant branch (SGB) stars is somehow the
most uncertain and arbitrary. In fact, no sharp separation is
theoretically expected between the cluster MS-TO/SGB and the BSS
sequence, which merges into the former with no discontinuity. Hence,
as a matter of fact, the BSS sequence must be truncated at a given
edge.  In doing this, we adopt a conservative criterion that naturally
takes into account the size of the photometric errors ($\sigma$) at
the required level of magnitude for the specific program cluster: we
define the lower boundary of the BSS selection box at $\sim
4$-$5\sigma$ from the magnitude and color distribution of the
MS-TO/SGB stars.  The red boundary of the selection box is generally
assumed to follow a vertical line aimed at excluding spurious objects
(generally originated by photometric blends) populating the ``plume''
observed just above the MS-TO. This feature is visible in many CMDs,
especially in central regions of high-density clusters, where
photometric blends are more probable.  Finally, the bright edge of the
BSS selection box is needed to distinguish very luminous BSSs from
stars populating the blue portion of the Horizontal Branch (HB). One
or two segments are required depending on the HB morphology and the
adopted combination of filters. It is important to emphasize that the
inclusion or exclusion of a few ``border'' objects usually has no or
negligible impact on the results.

To determine $A^+_{rh}$, the radial distribution of BSSs needs to be
compared with that of a REF population of normal (hence, lighter)
cluster stars tracing the overall density profile of the
system. According to the approach described in \citet{ferraro+18a}, we
choose to use as REF the MS stars around the MS-TO level.  Indeed,
this is the ideal REF population, as it includes several hundreds to
thousands of stars, and therefore is negligibly affected by
statistical fluctuations, thus allowing to maximize the accuracy of
the measure of $A^+_{rh}$.  Given the instrinsically large statistics
of this sample, the exclusion or exclusion of a few objects is even
less important than in the case of BSSs. Moreover, since the aim is to
trace the radial distribution of ``normal'' cluster stars, small
differences in the shape of the selection boxes are irrelevant. We
therefore traced the REF selection boxes with the aim to select the
bulk of stars in the MS-TO region.

Figures \ref{bss1640} and \ref{bss32} show the adopted selection
boxes drawn according to the scheme illustrated above, and the
resulting BSS and REF populations in the three program clusters. 

\subsection{Determination of the $A^+_{rh}$ parameter}
\label{sec:apiu}
Based on the evidence that internal dynamical processes make heavier
stars progressively sinking toward the center of a cluster more
rapidly than the less massive ones, the level of central segregation
of BSSs with respect to REF stars
is a powerful indicator of the stage of dynamical evolution reached by
the stellar system.  Hence, as discussed in the Introduction, the
$A^+_{rh}$ parameter has been specifically defined to quantify the
level of BSS central segregation in star clusters: it is the area
enclosed between the cumulative radial distribution of BSSs and that
of a REF population \citep{alessandrini+16}.  To allow the comparison
among stellar systems of different intrinsic sizes, the parameter is
built by using only the stars included within a fixed physical
distance from the cluster center (the half-mass radius).  Hence, in
building the cumulative radial distributions, the stellar distances
from the cluster center are normalized to $r_h$, and they are
expressed in logarithmic units to maximize the sensitivity of the
parameter to the innermost regions, where the efficiency of dynamical
friction is the highest (see equation 1 in \citealp{lanzoni+16}).
Moreover, to further enhance the sensitivity of $A^+_{rh}$ to internal
dynamical effects, \citet{ferraro+18a} suggested to consider only the
most massive tail of the BSS distribution, by selecting the stars
brighter than the MS-TO point in the UV CMD, in particular the BSSs
with $m_{\rm F275W}^*<-1$, where $m_{\rm F275W}^*$ is the F275W
magnitude normalized to the luminosity of the MS-TO of the host
cluster (i.e., by construction, the MS-TO has magnitude $m_{\rm
  F275})^*=0$). Although caution is needed in deriving BSS masses from
their luminosity (see \citealt{geller+11}), the brightest BSSs are
expected to be more massive than the fainter ones. In particular, the
adopted threshold in magnitude has been set to select BSSs that are
approximately $\Delta m\sim 0.2 M_\odot$ more massive than the MS-TO
stars.
 
To determine the $A^+_{rh}$ parameter in the three program clusters,
we followed the approach described above.  First, for the mass
selection of BSSs in the UV CMD of NGC 3201, we strictly followed
\citet{ferraro+18a}, i.e., we adopted $m_{\rm F275W}^*<-1$ (see the
dashed line in the bottom-left panel of Figure \ref{bbss}).  In all the
other cases, when only optical CMDs are available, we used theoretical
evolutionary tracks from the BaSTI database
\citep{pietrinferni+06,pietrinferni+21} to match the MS-TO region of
each system.  Since all the three clusters are very old stellar
systems, with ages of approximately 12-13 Gyr \citep{pallanca+21,
  deras+23}, we selected the 12.5 Gyr BaSTI isochrones with the
appropriate metallicity. In particular, considering that Galactic GCs
typically show $\alpha$-element enhancements of the order of
[$\alpha$/Fe]$\sim+0.3$, we adopted the Basti
$\alpha$-enhanced model with $Z=1\times 10^{-3}$ for NGC 3201 (having
[Fe/H]$\sim -1.6$; \citealt{harris+96}, and with $Z=8\times 10^{-3}$
for the two metal-rich systems (NGC 6440 and NGC 6316), where
[Fe/H]$\simeq -0.5/-0.6$ \citealt{harris+96,origlia+97,origlia+08}.
The adopted isochrones well reproduce the MS-TO region of the three
clusters (see the rightmost red lines in each panel of Figure
\ref{bbss}), and provided us with the mass of the stars evolving at
the MS-TO level: $\sim 0.8 M_\odot$ in the metal-intermediate system
NGC 3201, and $\sim 0.9 M_\odot$ in NGC 6440 and NGC 6316.  Hence, the
threshold adopted to select the bright portion of the BSS distribution
(dashed lines in Figure \ref{bbss}) has been set by considering the
location of the $1.0 M_\odot$ evolutionary track for NGC 3201, and the
$1.1 M_\odot$ model for NGC 6440 and NGC 6316 (leftmost red lines in
each panel of the figure).
The final samples of selected BSSs are shown as blue circles in Figure
\ref{bbss} and count 50, 44, and 38 objects in NGC 3201, NGC 6440, and
NGC 6316, respectively. Their normalized cumulative radial
distribution is plotted as a blue line in Figure \ref{apiu}, together
with that of the REF population, which is shown in red. The area of
the region included between the two cumulative distributions is shaded
in grey and corresponds to the value of $A^+_{rh}$ determined in each
cluster. It ranges from 0.19 in NGC 3201, up to 0.3 in NGC 6440, thus
indicating different levels of BSS segregation, corresponding to
different levels of internal dynamical evolution.  Following
\citet{ferraro+18a}, the errors on $A^+_{rh}$ have been estimated with
a jackknife bootstrapping technique \citep{lupton93}. In the case of
NGC 6316, for which a PM-based membership selection is not feasible
yet, we also evaluated the impact on $A^+_{rh}$ induced by the
potential contamination of Galactic field stars to the adopted BSS and
REF samples. To this end, we used the Gaia catalog between 500 and
$600\arcsec$ from the center (i.e., well beyond the cluster tidal
radius, $r_t=345\arcsec$; \citealp{deras+23}) to count the number of
field stars that fall within the two selection boxes \citep[see,
  e.g.,][]{dalessandro+19}].  By taking into account the areas covered
  by this field sample, and the area included within the cluster
  half-mass radius, we estimate that a total of 1 BSS (out of 38) and
  9 REF stars (out of 3707) could be field contaminants. We thus
  performed hundreds of random subtractions (in two separated radial
  bins) of 1 and 9 stars from the two respective samples, each time
  re-computing the value of $A^+_{rh}$. The result is that field
  contamination is totally negligible in this cluster.  The obtained
  values of $A^+_{rh}$ are listed in Table \ref{tab1}, together with
  those of the half-mass radius, $r_h$.

\subsection{Potential biases and uncertainties in the determination of the $A^+_{rh}$ parameter}
\label{sec:error}
As discussed in previous papers (e.g., \citealt{lanzoni+16,
  ferraro+18a, ferraro+19}), the main sources of possible biases in
the determination of the $A^+_{rh}$ parameter are photometric
incompleteness of the samples, and severe contamination from field
stars.  Being stronger in the innermost cluster regions, the effect of
incompleteness is to preferentially miss the most central objects.
Since field stars essentially have a uniform distribution within the
small sky area enclosed by a circle of radius $r_h$, the effect of
field contamination is to ``dilute'' the radial distribution.  Hence,
in general both these biases tend to underestimate the value of
$A^+_{rh}$.  This is the reason why we always adopt methodologies
specifically designed to reduce, and possibly eliminate, their impact.
Indeed, we always follow the approach of sampling the most crowded
central region of the investigated star clusters with high-resolution
HST images (acquired in UV or optical filters).  Moreover, considering
only the brightest portion of the BSS sequence for the measure of
$A^+_{rh}$ not only maximizes the central sedimentation effect (see
Section \ref{sec:apiu}), but it also reduces the impact of
incompleteness. This is also supported by the artificial star
experiments, which confirm the large level ($>90\%$) of photometric
completeness for the adoped BSS samples.  To minimize/avoid the
potential bias introduced by field star contamination, in most of the
cases we used PM-selected samples. When PMs are not available (e.g.,
for NGC 6316 and the GCs in the LMC), a statistical decontamination of
the samples has been performed by evaluating the number of potential
field stars, running hundreds of random subtractions of them from the
selected BSS populations, and thus including this effect in the
overall uncertainty of the measure.
 
Once incompleteness and field contamination are under control, the
primary source of uncertainty on $A^+_{rh}$ remains the relatively
small-number statistics of the BSS sample. As mentioned above, to
estimate the errors on $A^+_{rh}$ we adopted a jackknife bootstrapping
technique \citep{lupton93}: operatively, for a sample of $N$ BSSs, we
determined the value of $A^+_{rh}$ $N$ times by using samples of
$(N-1)$ BSSs obtained by excluding, each time, one different
star. Thus, the procedure yields $N$ estimates of the parameter and
the final uncertainty on $A^+_{rh}$ is obtained as
$\sigma_{A+}=\sigma_{\rm distr} \times \sqrt{(N-1)}$, where $\sigma_{\rm
  distr}$ is the standard deviation of the $A^+_{rh}$ distribution
derived from the $N$ realizations.

\section{Discussion}
\label{sec:discussion}
\begin{figure}[ht!]
\centering
\includegraphics[width=8cm, height=8cm]{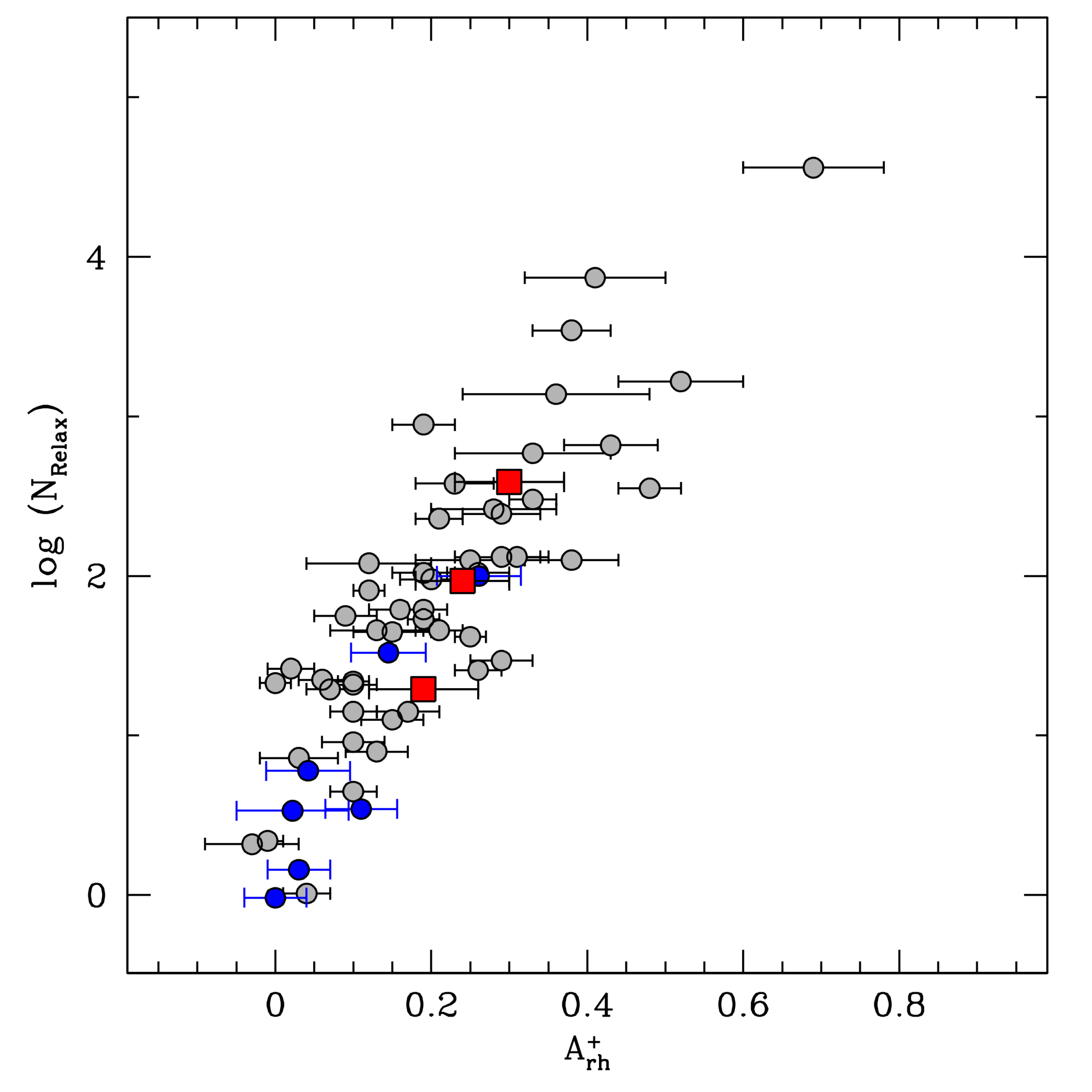}
\centering
\caption{Correlation between the BSS segregation level (measured by
  the $A^+_{rh}$ parameter) and the number of current central
  relaxation times occurred since cluster formation ($N_{\rm relax}$)
  in the sample of star clusters analyzed so far: the 3 systems
  studied here are highlighted with large red squares, the 48 Galactic
  GCs discussed in \citet{ferraro+18a} and the cluster (NGC 6256)
  studied in \citet{cadelano+22} are plotted as grey circles, while
  the 5 LMC and the 2 SMC clusters presented in \citet{ferraro+19} and
  \citet{dresbach+22}, respectively, are marked with blue
  circles.}
\label{apiu_nrelax}
\end{figure}

The values of $A^+_{rh}$ here determined for the three program
clusters can now be compared with those obtained in previous studies
through a similar methodology.  Following \citet{ferraro+18a}, the
comparison is worth to be done in a diagram relating the values of
$A^+_{rh}$ with the number of present-day central relaxation times
suffered by each system since the epoch of its formation ($N_{\rm
  relax}$).  The value of $N_{\rm relax}$ can be derived by simply
dividing the age of the cluster\footnote{Accordingly to
\citealt{ferraro+18a} we adopted 12 Gyr as average age for GGCs, see
the compilation by \citealt{forbes+10} } by its central relaxation
time, estimated from the well-known expression
\citep{spitzer87,djo93}:
\begin{equation}
t_{rc} = 8.338 \times 10^6\times  \ln(0.4 N_*)^{-1} (\rho_0)^{1/2} (m_*)^{-1}
r_c^3,
\label{eq1}
\end{equation}
where $N_*$ is the total number of stars computed as the ratio between
the total cluster mass ( $M_{\rm tot}$) and the average stellar mass
(here we adopt $m_*=0.3 M_\odot$), $\rho_0$ is the central mass
density in units of $M_\odot$ pc$^{-3}$, determined through eq.(7) in
\citet{djo93}, and $r_c$ is the core radius in pc. In turn, $M_{\rm
  tot}$ is obtained from the product between the absolute $V-$band
magnitude of the cluster ($M_V$) listed in the \citet{harris+96}
catalog and the $V-$band mass-to-light ratio appropriate for old
stellar systems ($M/L_V=2$; e.g., \citealp{maraston98}).  Figure
\ref{apiu_nrelax} compares the values of $\log(N_{\rm relax})$
\emph{versus} $A^+_{rh}$ obtained for the three clusters analyzed here
(large red squares), with those previously measured in a consistent
way for other systems, namely, the 48 Galactic GCs presented in
\citet{ferraro+18a} and NGC 6256 from \citet[][grey
  circles]{cadelano+22}, 5 old clusters in the LMC and 2 young SMC
clusters discussed, respectively, in \citet{ferraro+19} and
\citet[][blue circles]{dresbach+22}.  Considering the three systems
studied here, a grand total of 59 stellar clusters has been
investigated so far in different environments.  As apparent from
Figure \ref{apiu_nrelax}, they draw a well defined relation between
$A^+_{rh}$ and $N_{\rm relax}$, thus proving and consolidating the use
of the dynamical clock as a powerful method to track the dynamical
evolution of stellar systems, and the use of $A^+_{rh}$ as a sensitive
clock-hand to rank star clusters in terms of their dynamical age,
irrespective of the host environment.

The tight relation shown in Figure \ref{apiu_nrelax} indicates that
$A^+_{rh}$ and $N_{\rm relax}$ both provide a measure of the dynamical
ageing of stellar systems. However, it is worth adding a few
considerations about the two indicatoris.  The dynamical evolution of
a cluster is driven by the complex combination of effects associated
to a variety of internal and external properties and dynamical
processes (e.g., two-body relaxation, interactions with the Galactic
tidal field along the cluster orbit, initial cluster structural and
kinematic properties and its stellar content; see e.g. Heggie \& Hut
2003).  The value of $t_{rc}$ provides a measure of the present-day
central relaxation timescale and it is based on a simple analytical
expression derived under the assumption of spherical symmetry, and
isotropic and non-rotating internal kinematics. However, recent
observational investigations \citep[e.g.][]{fabricius+14, watkins+15,
  bellini+17, kamann+18, ferraro+18b, lanzoni+18a, lanzoni+18b,
  leanza+22} have revealed that some GCs are characterized by internal
rotation and velocity anisotropy, and several theoretical studies have
shown that these kinematic properties have a significant effect on all
the aspects of the dynamical evolution of GCs \citep[see,
  e.g.][]{kim+04, hong+13, tiongco+17, breen+17, pavlik+21,
  pavlik+22a, pavlik+22b, kamlah+22, livernois+22}.  In addition,
depending on the central density, core radius, and total mass, the
value of $t_{rc}$ computed through eq. (1) varies during cluster
dynamical evolution. Hence, the present-day value of $t_{rc}$ may fail
to provide a complete picture of the past evolutionary history of a
system.  In fact, by construction, two clusters with similar
present-day structural properties would share the same value of
$t_{rc}$, even though they experienced different dynamical
evolutionary histories.  Hence, using this value as an estimate of the
systems' dynamical age may not capture differences in their past
evolution.  On the observational side it is also worth mentioning the
large uncertainties in the derivation of reliable estimates of
$t_{rc}$ in the case of post core collapse (PCC) clusters, where the
presence of a inner cusp in the star density/surface brightness
profile allows no proper fit with the King model family, and therefore
prevents any reasonable measure of $r_c$ (and possibly invalidates its
meaning itself). Indeed, the difficulty in determining the structural
parameters of PCC clusters possibly is at the origin of the large
spread of $t_{rc}$ (hence, $N_{\rm relax}$) values in the most
advanced stages of dynamical evolution (see the top-right corner of
Figure 7).  On the other hand, by leveraging the observed spatial
concentration of the BSS populations, $A^+_{rh}$ provides a direct
empirical measure of the degree of mass segregation developed during
the \emph{entire} cluster's evolution.  In addition, such
observational measure does not rely on simplifying assumptions and
approximations, and represents a more direct indicator of the
cluster's dynamical history than the present-day value of $t_{rc}$.

A few cases provides interesting examples of how the $A^+_{rh}$
parameter may reveal a more detailed picture of the dynamical history
of globular clusters.  NGC 4590 and M3 (NGC 5272 have the same
relaxation time ($\sim 4.6 \times10^8$ yr), but very distinct values
of $A^+_{rh}$: 0.02 and 0.26, respectively.  The latter clearly shows
that M3 is dynamically much older than NGC 4590, consistently with its
more compact structure and higher central density.  Similarly, the
dynamical clock indicates that, in spite of the same relaxation time
($t_{rc}\simeq 3.1\times 10^7$ yr), NGC 6440 ($A^+_{rh}=0.30$) is more
dynamically evolved than NGC 6535 ($A^+_{rh}=0.23$), which is a
significantly less compact and less massive cluster.  Note that the
high dynamical activity in the core of NGC 6440 is also testified by
the presence of a significant population of millisecond pulsars
(MSPs): 8 MSPs have been found so far in this stellar system (see
\citealt{freire+08, vleeschower+22}). More intriguingly, NGC 6440 is
one of the three Galactic GCs (beside M28 and NGC 2808) hosting a
so-called accreting millisecond X-ray pulsar, a subgroup of transient
low-mass X-ray binaries that show, during outbursts, X-ray pulsations
from a rapidly rotating neutron star (see \citealt{sanna+16,
  cadelano+17a}). The object in NGC 6440 is characterized by an ongoing
mass transfer process that, according to the currently accepted
recycling formation scenario \citep{bhatta+91}, as soon as the radio
signal is reactivated will yield to the appearance of a new-born MSP
(see \citealt{ferraro+15}) in the radio band. This, together with the
large BSS sedimentation level testified by the measured value of
$A^+_{rh}$, can be considered as a clear signature of an intense
dynamical activity occurring in the core of a cluster on the verge of
CC.  A deeper investigation of these cases can provide illuminating
details on the processes that contribute to determine the dynamical
aging of star clusters.

As quoted above, Figure \ref{apiu_nrelax} also includes 7 clusters in
the LMC and SMC with old and intermediate chronological ages,
respectively, which have been found to span values of $A^+_{rh}$
between 0 and 0.26 (blue circles in the figure).  Extending the
measure to the PCC clusters of the LMC is now urged to probe, also in
an extra-Galactic context, the entire range of dynamical ages sampled
by the $A^+_{rh}-N_{\rm relax}$. However, the nice way in which the
LMC and SMC clusters merge into the Milky Way relation suggests that
this holds also in eternal galaxies.  In particular, the old LMC
clusters NGC 2210, NGC 2257 and Hodge 11 turn out to essentialy share
the same dynamical ages (and similar structural parameters) of the
Milky Way GCs M92 (NGC 6341), NGC 5466, and NGC 6101,
respectively.  
 
In addition, the null value of $A^+_{rh}$ (indicating that BSSs have
yet to start their central sedimentation) that has been measured in
the intermediate-young SMC clusters NGC 339 and NGC 419 is well
comparable to that observed in three old GCs in the Milky Way (namely,
$\omega$Centauri, NGC 2419 and Palomar 14), and in two old GCs in the
LMC (NGC 1841 and Hodge 11). This evidence strongly suggests that a
lack of BSS segregation (i.e., a BSS radial distribution
indistinguishable from that of lighter stars) can be reasonably read
as the {\it initial} condition of any cluster, before internal
dynamical processes start to significantly modify the spatial
distribution of stellar masses and the overall structure of the
system. Hence, the large core radius measured in various old clusters
of the LMC can just correspond to the initial conditions at the moment
of formation \citep[see][]{ferraro+19}, instead of being the result of
a core expansion due to the action of a binary black hole population
(as suggested by \citealt{mackey+08}).

\begin{figure}[ht!]
\centering
\includegraphics[width=8cm, height=8cm]{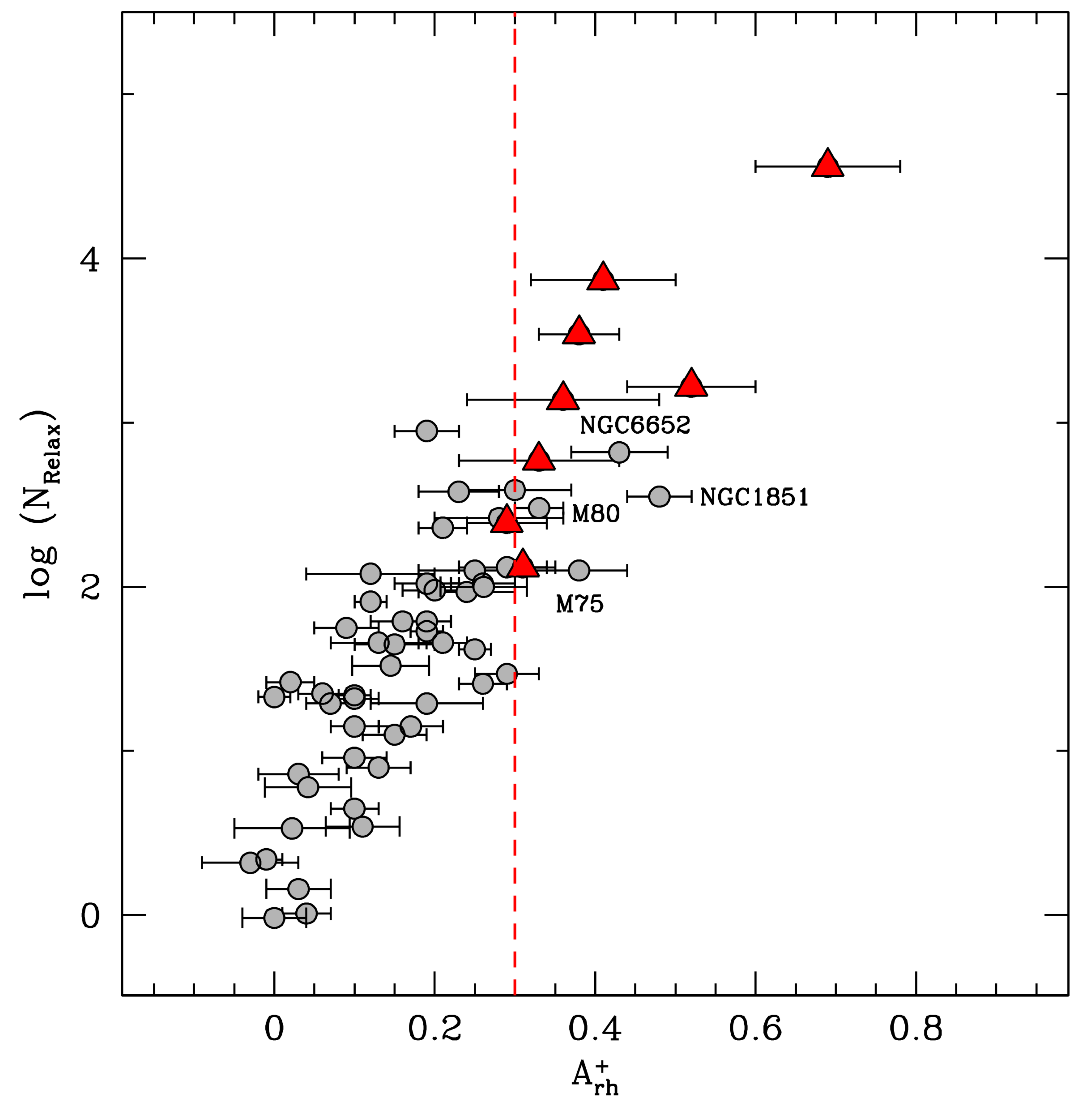}
\centering
\caption{Same as in Figure \ref{apiu_nrelax}, with the position of the
  8 confirmed PCC clusters highlighted with large red triangles. The
  vertical dashed line at $A^+_{rh}=0.30$ likely flags the reference
  value of $A^+_{rh}$ for the occurrence of CC. The four clusters with
  $A^+_{rh}>0.30$, but no clear evidence of CC based on the shape of
  their density profile are labelled.}
\label{fig:pcc}
\end{figure}

Interestingly, the 7 Galactic GCs in the surveyed sample that are
classified as PCC systems in the \citet{harris+96} catalog (namely,
M15, M30, M70, NGC 6397, NGC 6624, NGC 6256 and NGC 6752) all have
$A^+_{rh}\ge 0.29$, and the same holds for NGC 362 that is a suspected
PCC system in the \citet{harris+96} catalog and confirmed so in
\citet{dalessandro+13}: they are plotted as large red triangles in
Figure \ref{fig:pcc}. This evidence provides a series of additional
considerations on the potential information provided by the dynamical
clock.  In fact $A^+_{rh}\sim 0.30$ (dashed line in the figure) can be
considered a sort of reference value for the CC event, meaning that
the proximity of $A^+_{rh}$ to this value likely is an indication of
the imminence of CC.  In this respect, GCs as NGC 6440, NGC 6229, and
47 Tucanae (with $A^+_{rh}=0.29-0.30$) should be very close to CC.  On
the other hand, very intriguing cases are the four clusters (namely
M80, M75, NGC 6652 and NGC 1851) showing $A^+_{rh}>0.33$ and no
evidence (at least known so far) of a steep cusp in the innermost
portion of the star density/brightness profile, which is considered to
be the typical signature of CC.  These clusters are surely worth of
deeper investigations to confirm that they show no signatures of CC.
  If so, the large values $A^+_{rh}$ could be interpreted as the
  manifestation of a population of collisional and mass-transfer BS
  generated into the core by an increased rate of binary interactions
  that are contributing to delay CC. In this case, also other signatures 
of this activity should be detectable since compact binaries forming and 
hardening in the core can manifest themselves as interacting binaries. 
Intriguingly, a high rate of ongoing
dynamical interactions in the core of NGC 1851 and NGC 6652 is
suggested by the presence of recently formed MSPs, possibly generated
by exchange interactions (\citealp{ridolfi+22}, and Chen et al. 2023
in preparation, respectively).  In turn, this would allow us to
extract additional information from the sedimentation level of BSSs.
In fact, if no signatures of CC are detected, the large values of
$A^+_{rh}$ measured in these clusters would be interpreted as if they
were due to a collisional and mass-transfer population of BSSs
generated into the core by the increased rate of binary interactions
that are contributing to delay CC.  On the other hand, the large range
of values of $A^+_{rh}$ found for PCC clusters (from $\sim 0.3$ to
$\sim 0.7$) suggests that an intense and progressive growth of the BSS
population in the cluster center takes place also during the PCC
stage, again possibly tracing an increased formation of these stars
due to stellar collisions and enhanced mass transfer activity in
(hard) binaries.  Interestingly enough, a possible confirmation of
this activity arrives from the analysis of the rotational velocity of
BSSs in PCC clusters (see Figure 3 in \cite{ferraro+23}). In fact, the
fraction of fast spinning BSSs (which is a signature of recent
formation; \citealp{sills+05, leiner+18}) shows a (mild) increase with
$A^+_{rh}$, which might indicate that the dynamical activity occurring
during the recurrent re-contractions of the cluster core in the PCC
evolutionary stage could favor the formation of these stars.

\begin{deluxetable*}{lrcccl}
\tablecaption{$A^+_{rh}$ values determined for  the program clusters}
\tablewidth{0pt}
\renewcommand{\arraystretch}{1.2}
\tablehead{
\colhead{ Cluster } & \colhead{ [Fe/H] }  & \colhead{$r_{h}$}  & \colhead{$\lg(t_{rc})$} & \colhead{$A^+_{rh}$} & \colhead{Reference}}
\startdata
NGC 6440 &  $-0.5$  &  $50\arcsec$  & 7.49 &  $0.30\pm0.07$ & \colhead{Pallanca et al., 2021} \\
NGC 6316 &  $\sim -0.6$  & $40\arcsec$  & 8.11 &  $0.24\pm0.06$ & \colhead{Deras et al., 2023}  \\
NGC 3201 &  $-1.6$  & $272\arcsec$  & 8.79 &  $0.19\pm0.07$  & \colhead{Lanzoni et al. 2023 (in preparation)}\\
\enddata
\end{deluxetable*}
\label{tab1}

All the considerations above support the importance of the dynamical
clock and further sustain its potential use in the near future as a
powerful indicator of dynamical evolution.  In fact, the upcoming
generation of telescopes will open the exploration of resolved
populations in star clusters in external galaxies, allowing a
straightforward selection of BSSs and the study of their radial
distribution. Indeed, the observational capabilities of the JWST
  already extended the exploration of resolved stellar populations
  below the MS-TO level in the GC systems of the entire Local Group,
  including the populous systems in the Andromeda galaxy, in M33 and
  in many nearby dwarf galaxies. This exploration will be further
  pushed forward by the Multi-AO Imaging Camera for Deep Observations
  (MICADO) that will be mounted at the ESO Extremely Large Telescope
  (ELT, a telescope with a 39-metre primary mirror) and will provide a
  spatial resolution a factor of 6 better than the JWST.  Indeed, for
these distant systems,
the $A^+_{rh}$ parameter will effectively be the easiest (if not the
only) diagnostic available to study their dynamical stage. In turn,
this will promote a variety of science cases, allowing, e.g., the
appropriate selection of dynamically-old clusters on the verge of CC,
the identification of dynamically-young systems where to study the
initial conditions of their (possibly multiple) populations, up to the
study of the effects that the interaction with the parent galaxy may
have on the dynamical aging of star clusters.


\vskip1truecm
This work is part of the project {\it Cosmic-Lab} ({\it Globular Clusters as Cosmic Laboratories}) at the Physics and
Astronomy Department "A. Righi" of the Bologna University
(http://www.cosmic-lab.eu/ Cosmic-Lab/Home.html). The research was
funded by the MIUR throughout the PRIN-2017 grant awarded to the
project {\it Light-on-Dark} (PI:Ferraro) through contract PRIN-2017K7REXT.


\newpage

\bibliographystyle{aasjournal}



\end{document}